# Coping with space neophobia in *Drosophila melanogaster*: The asymmetric dynamics of crossing a doorway to the untrodden

Shay Cohen[1,3], Yoav Benjamini[2,3], Ilan Golani[1,3]

1. Department of Zoology, Tel Aviv University, Tel Aviv, Israel.

2. Department of Statistics and OR, Tel Aviv University, Tel Aviv, Israel,

3. Sagol School of Neuroscience, Tel Aviv University, Tel Aviv, Israel

Abstract

Insects exhibit remarkable cognitive skills in the field and several cognitive abilities have been demonstrated in Drosophila in the laboratory. By devising an ethologically relevant experimental setup that also allows comparison of behavior across remote taxonomic groups we sought to reduce the gap between the field and the laboratory, and reveal as yet undiscovered ethological phenomena within a wider phylogenetic perspective. We tracked individual flies that eclosed in a small (45mm) arena containing a piece of fruit, connected to a larger (130mm) arena by a wide (5mm) doorway. Using this setup we show that the widely open doorway initially functions as a barrier: the likelihood of entering the large arena increases gradually, requiring repeated approaches to the doorway, and even after entering the flies immediately return. Gradually the flies acquire the option to avoid returning, spending more relative time and performing relatively longer excursions in the large arena. The entire process may take up three successive days. This behavior constitutes coping with space neophobia, the avoidance of untrodden space. It appears to be the same as the neophobic doorway-crossing reported in mouse models of anxiety. In both mice and flies the moment-to-moment developmental dynamics of transition between trodden and untrodden terrain appear to be the same, and in mice it is taken to imply memory and, therefore, cognition. Recent claims have been made for a deep homology between the arthropod central complex and the vertebrate basal ganglia, two structures involved in navigation. The shared dynamics of space occupancy in flies and mice might indicate the existence of cognitive exploration also in the flies or else a convergent structure exhibiting the same developmental dynamics.

Introduction
The remarkable cognitive skills of insects have attracted the attention of naturalists for decades, including the complex routines exhibited by wasps attending to their offspring [1], the intricate navigational capacities exposed in ants [2], and the use of landmarks during return flights to a food source in honey bees [3]. Several cognitive abilities in *Drosophila melanogaster* behavior have been demonstrated in the laboratory. In a y-maze test, flies exhibited memory after being trained to avoid an arm coupled with electric shock, and were subsequently shown to identify and remember visual features such as odor and colored light [4], as well as size, color, and contour orientation [5] In a thermal-visual arena inspired by the Morris Water Maze [6] and the Heat Maze [7, 8], flies had to find and then exhibit memory of the location of a cool tile, based on remote spatial cues, in an otherwise hot environment [9]; male fly courtship behavior was shown to be modified by prior sexual experience [10]; and in a flight simulator tethered flies controlling the horizontal rotations of an arena exhibited memory templates [11], and discrimination and generalization learning [12].

Our aim here was to determine whether *Drosophila melanogaster* distinguishes between visited and not visited (or less visited), areas of an environment; whether it behaves in a neophobic fashion and, if so, to describe the dynamics of the process and to quantify it. To achieve these aims it was necessary i) to design an experimental environment that would clearly distinguish between trodden and untrodden (unforeseen) parts of that environment: a gradual increase in activity in the untrodden part of the environment, and a concurrent gradual relative decrease in activity in the well-trodden part would suggest a change in relative preference for each of the two parts, differential recognition of each of them, spatial orientation, and choice between them; ii) to minimize coercion in order not to confound spontaneous and reactive behavior [13]; iii) to determine the time frame that would correspond to the phenomenon's intrinsic time scale; iv) to represent behavior as a product of a dynamic interaction between the animal and the environment; and v) to allow the operational significance of the behavior to emerge.

The experimental environment consisted of a small (45mm) arena with a piece of fruit in it, connected to a large (130mm) empty arena by a wide (5mm), permanently open doorway. In nature, male fruit-flies often eclose near or on fermenting fruit, feed on the yeast that grows on it [14, 15] fight with other males, and court and copulate with females, which then complete a life cycle by oviposition on the fruit. While the fruit serves as a naturalistic attractor and center around which flies are expected to perform ethologically relevant behavior, flies are clearly not likely to encounter doorways in their natural habitat. Confronting wild organisms with supernormal stimuli that enhance their species-specific behavior to an exaggerated degree is, however, a standard methodology used by ethologists [16]. In our setup the doorway enhances the contrast between trodden and untrodden terrain, amplifying the response to the contrast. Two other ethological features of our experiment are sparing the fly a forced introduction into the setup by allowing it to eclose in the vicinity of the fruit, and perform at its own pace, without stress, for the time frame of three days found by us in preliminary experiments to be necessary for the phenomenon to unfold. The wide open doorway leading to a large empty arena provides a necessary distinction between two connected spaces. A quantification of the two-way doorway-crossing reveals the gradually changing functional permeability of the doorway and the concurrent change in the behavior of the fly in each of the two arenas. This is brought about by the fly's use of what appears to be an operation that changes the doorway's functional permeability.

Results

The smoothed output of the fly's location time-series contains an average of more than 5.5 million data points that depict ca. 200 meters of walking distance covered during a single 3-day session. Figure 1 illustrates the paths traced by a selected fly in the course of 30 hours, just prior to the first entry into the large arena (1a); just after the first holding back (1b); and over the course of 60 hours (1c). To study the path's developmental dynamics we performed low level partitioning of the path into progression and lingering segments [17] and higher level partitioning into entries (into the large arena) and departures (back to the small arena; see Methods).

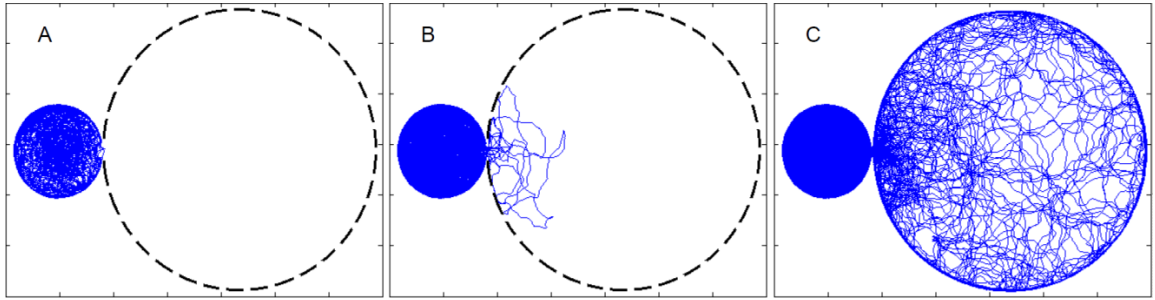

**Fig.1. The path traced by a fly in an experimental setup consisting of a small and a large arena connected by a 5 mm width doorway in the course of a session.** (a) after 29 hours, just before the first entry into the large arena, (b) after 36 hours, just before the first holding back, and (c) after 60 hours.

*The developmental dynamics of doorway crossing in fruit flies*

The flies' experimental setup comprised two arenas: a small one, in which the fly ecloses and where the fruit is located; and a large empty one connected to it by a doorway 5mm in width, about 5 times wider than the fly. We correspondingly divided the fly's path into "entries" into the large arena and "departures" back into the small arena: an entry was defined as a segment that starts when the fly enters the large arena and ends when it leaves it; a departure was defined as a segment that starts when the fly enters the small arena and ends when it leaves it.

Our preliminary observations showed that the flies did not tend to cross into the large arena and, when they did, the first entries were short while the later ones became longer over time. Therefore, we plotted the entries successively in order to obtain a view of the developmental dynamics of entries across a session (Video A1).

Unlike the clear buildup in the extent and complexity of paths illustrated in video A1, in the behavior of a selected fly, observation alone does not show a clear developmental trend in each of the flies; analysis of the data is necessary in order to determine whether trends exist.

Another phenomenon that we observed in preliminary examination of the behavior was that the fly frequently arrived at the doorway but did not cross it although it could have readily done so. However, as time passed, the fly appeared more likely to cross the doorway upon approaching it. In order to test this hypothesis quantitatively, we examined what happened each time the fly approached the doorway while still in the small arena. Visits to the doorway were defined by a circular area centered at the doorway (see Methods). Once the fly had visited the defined circular area we established whether it i) began to cross the doorway into the large arena (figure 2b); or ii) exited the circle without crossing the doorway (figure 2a).

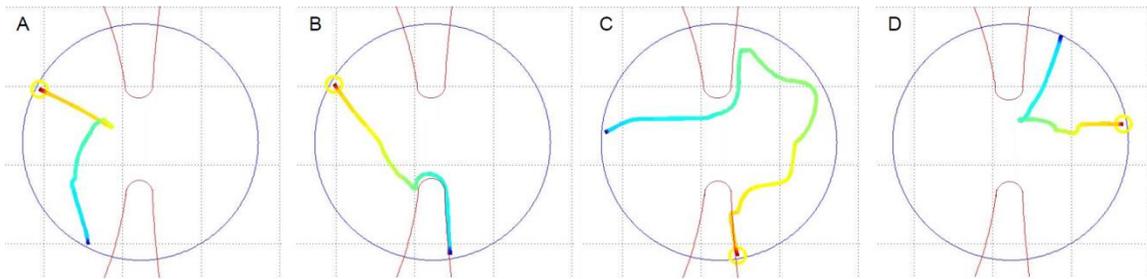

**Fig.2. The four behavioral options exhibited by a fly upon approaching the doorway.** A distance of 8.5mm or less from the doorway's center defines proximity to it. Path segments traced by the fly are colored from yellow (the beginning of the segment) to light blue (the end of the segment). The fly approaches the doorway (a) from the small arena but does not cross to the large arena, (b) from the small arena crossing into the large arena (cutting through), (c) from the large arena crossing into the small arena (d), from the large arena but does not cross to the small arena (holding back).

One approach to the sequence of visits was to partition it into a specific number of bins, e.g., 20, and to plot the percentage of crosses per bin (figure 3). As illustrated, the probability of crossing the doorway, reflected by the percentage of crosses per bin, tended to increase across the sequence of visits bins.

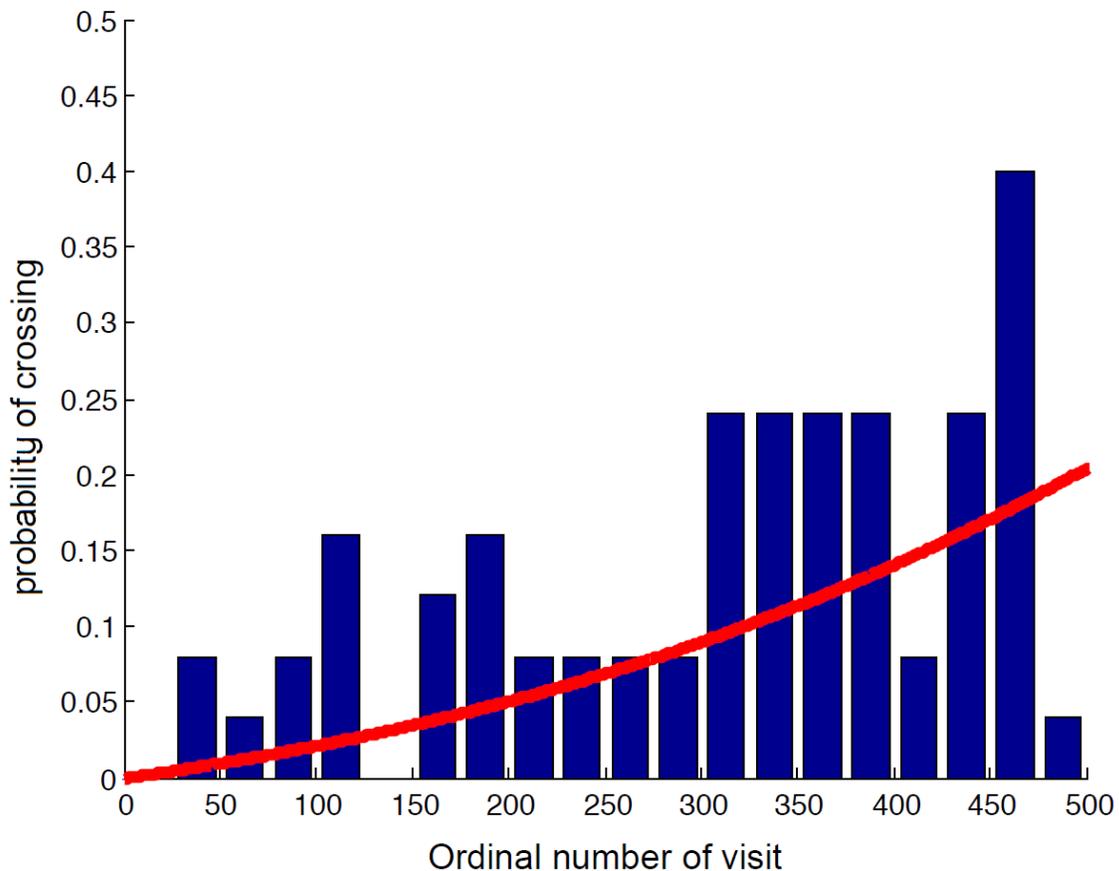

**Fig.3. The percentage of doorway crossing per bins of 20 visits to doorway area with a logistic regression plotted on top of it, in a selected fly**. Note the increase in the likelihood of cutting through across the session.

Moreover, a logistic curve superimposed on the bin plot displayed in figure 3, indicates that a reasonable model for the probability of crossing the doorway can be the logistic one (see Methods).

To avoid the arbitrariness associated with the number of bins and number per bin, as seen in the figure, we directly quantify the probability of crossing the doorway as a function of the visit number, separately for each fly. Using the fitted logistic curves for each of the 24 flies, figure 4 presents the *change* from its initial value of the probability of crossing from the trodden to the less trodden arena (note that the initial probability of crossing is not necessarily the same for all flies). As shown, 22 out of 24 flies showed an increase incrossing probability.

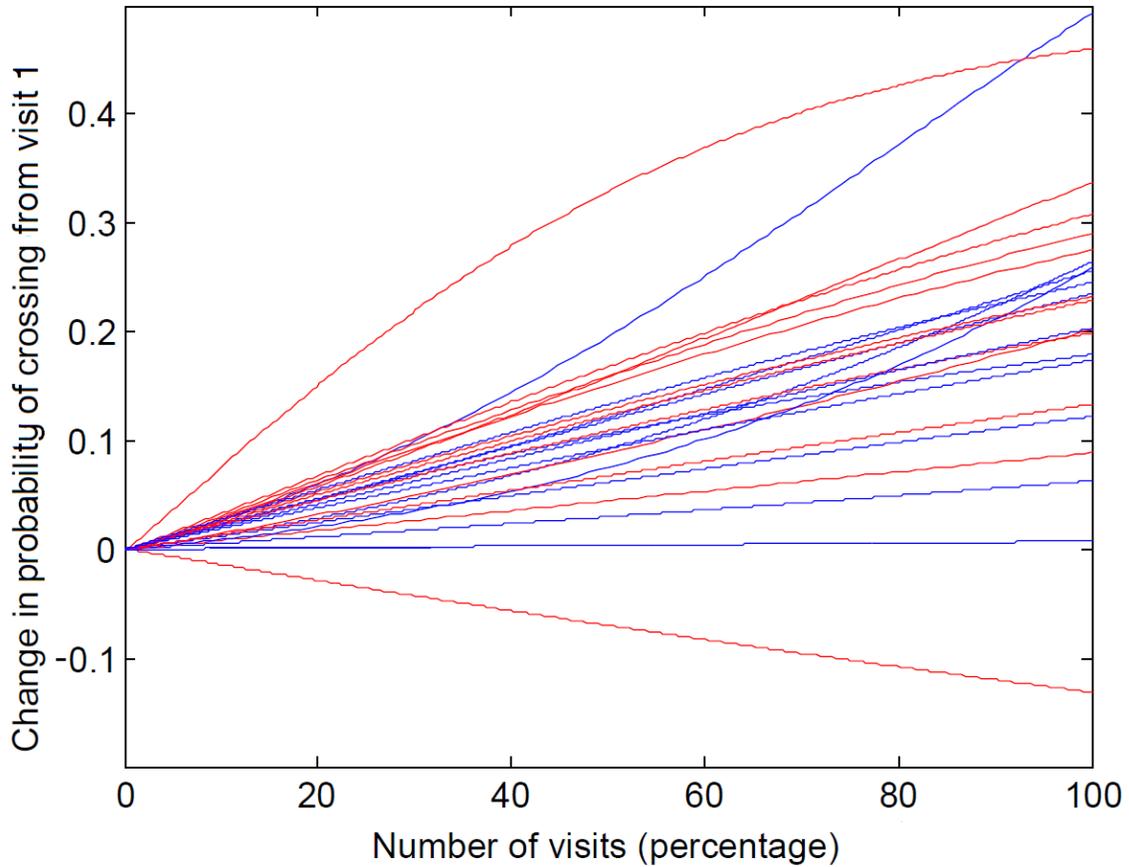

**Fig.4. Dependency of odds to cross from the well-trodden to the less trodden arena on visit number, in individual flies.** The X-axis represents the ordinal number of visit (percentage out of total amount of visits). The Y- axis represents the difference in the likelihood of crossing, across the session. Black lines signify males and red lines females.

The fitted logistic model per fly allows us to quantify the rate of the increase of probability of crossing the doorway. In particular, the logarithm of the odds of crossing the doorway (i.e logarithm of p/(1-p) ) is linear in visit number with the estimated slope denoted by β: the higher the positive beta is, the faster is the increase, and obviously a negative beta indicates decrease.

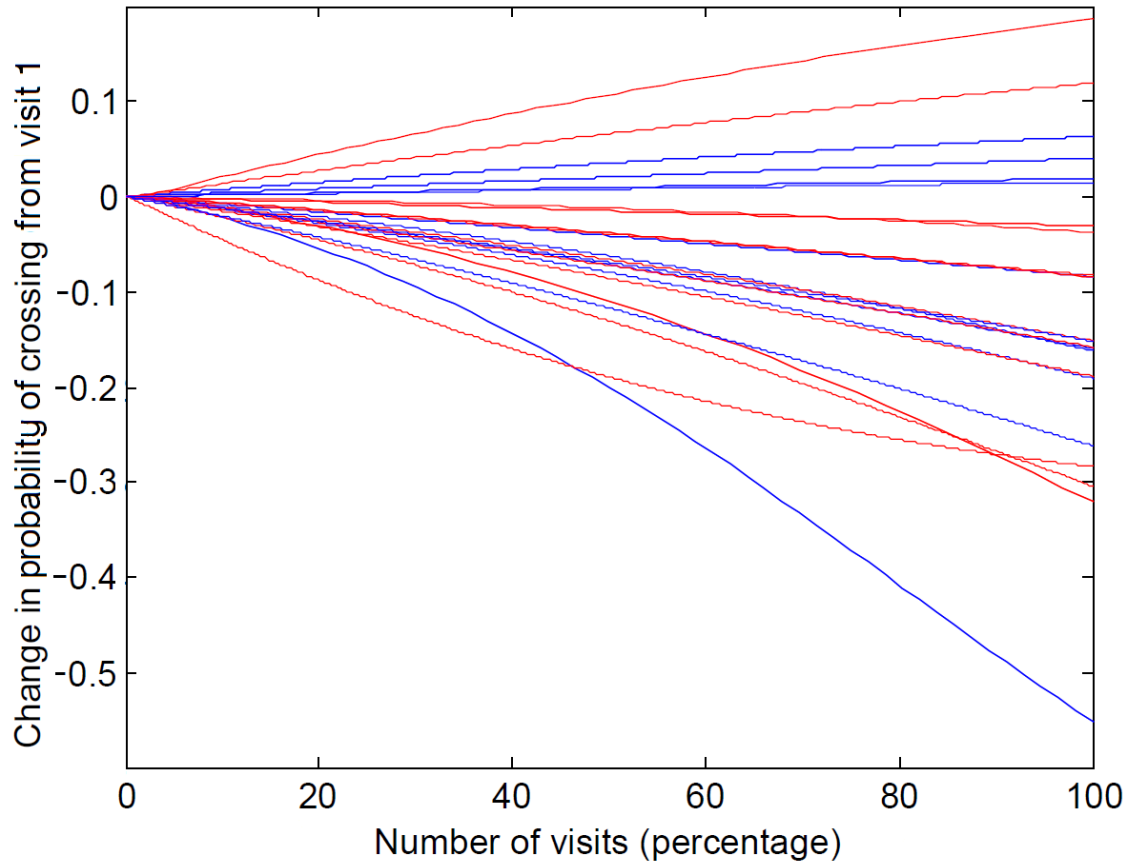

**Fig.5. Dependency of odds to cross from the less trodden to the well-trodden arena on visit number, in individual flies.** The X-axis represents the ordinal number of visit (percentage out of total amount of visits). The Y- axis represents the difference in the likelihood of crossing across the session. As shown, 17 out of 24 flies show a decrease in crossing likelihood. Black lines signify males and red lines females.

Twenty two of the 24 flies exhibited positive slopes, indicating increasing probability of door crossing with visit number at doorway ($p \leq 0.0001$, using sign test).

The most significant trend ($p<0.0001$) for the flies as a group was demonstrated for a radius of 8.5 mm (Figure 4; see methods).

In contrast, the probability of crossing in the opposite direction – from the large to the small arena (Figure 2c, 2d) tends to decrease, with 18 of the 24 flies having negative slopes ((p=0.0066, using the sign test). In summary, the more times a fly entered the pre-defined circular area, the higher its probability of crossing the doorway into the large arena and the lower chance of crossing the doorway back into the small arena (Figure 5).

The difference in slopes from small to large arena between males and females was p<0.126 and from large to small arena p<0.355 with 95% confidence interval (using Wilcoxon-Mann-Whitney). The difference is small and not statistically significant at 5%.

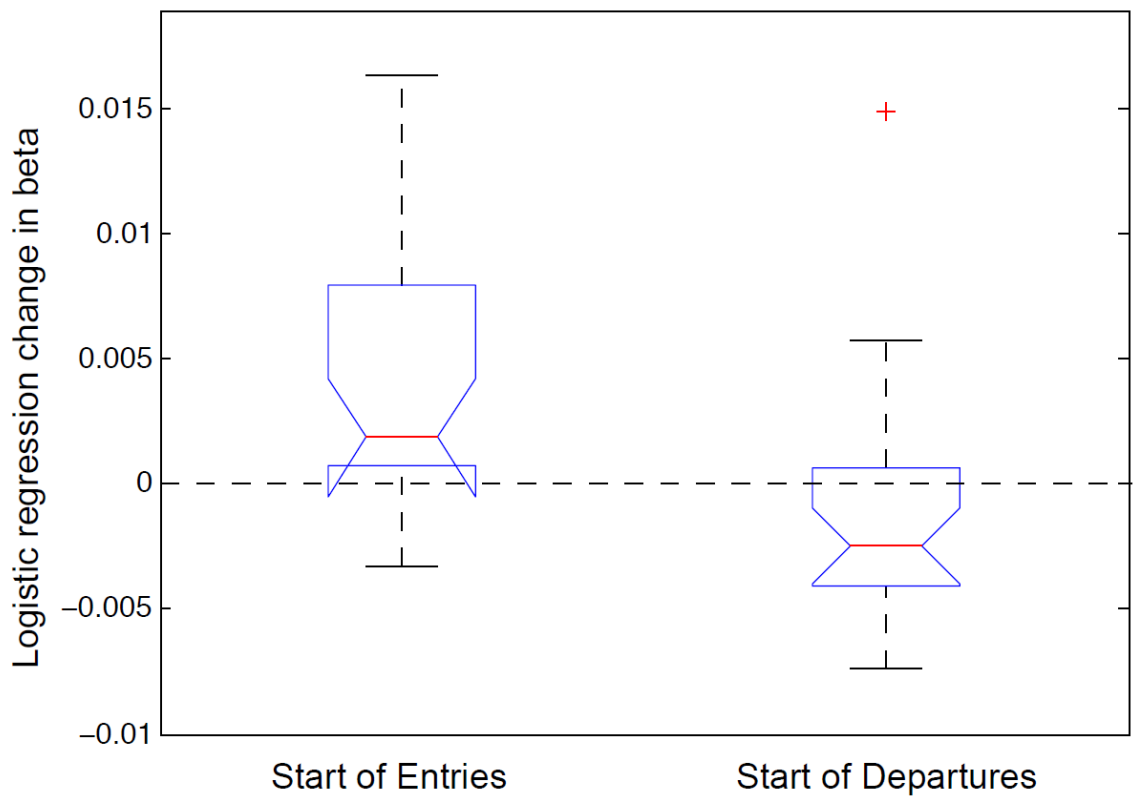

**Fig.6. Comparison between the developmental dynamics of doorway crossing from small to large arena (start of entries) and from large to small arena (Start of departures).** The comparison is between trends in the Beta of the logistic regression - the flies show an increase in the likelihood of crossing into the large arena (p<0.0001), and a decrease in the likelihood of crossing into the small arena (p=0.0066).

We have shown so far that the trends in doorway-crossing from small to large and from large to small arena were opposite, reducing the flies' bias towards the small arena. Our next question was whether the bias would also be expressed in a change in the relationship between the distances traveled and times spent in each of the two arenas. To answer these questions we divided the distance traveled in each entry by the distance traveled in the departure that preceded it, and examined whether there was a change across the sequence of ratios of distances, and whether this was negative or positive.

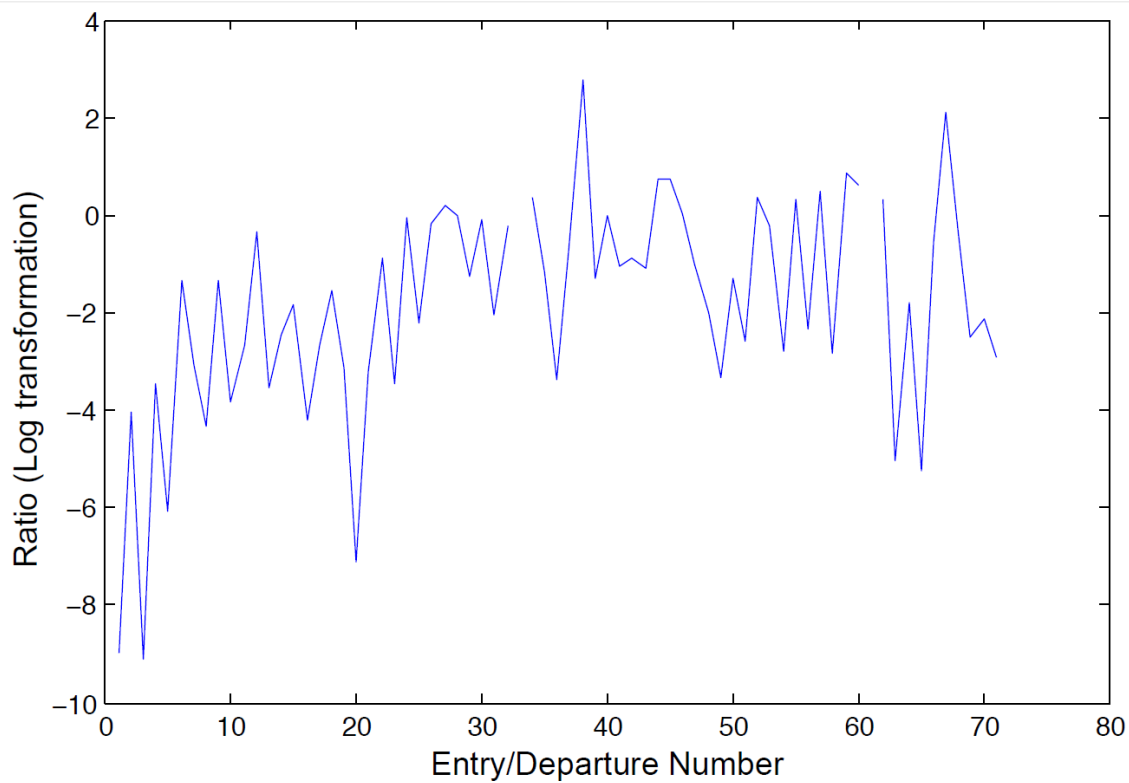

**Fig.7. The developmental dynamics of the ratio between distance traveled per entry and distance traveled in its successive departure.** As shown, in this specific fly, there is a gradual increase in the relative distance traveled in the large arena.

Plotting the sequence of logarithm of ratios of distances in a selected fly indicated a positive trend across most flies, implying a progressive increase in distance traveled in the large arena relative to distance traveled in the small arena (Fig.7). The trend was summarized per each fly by the Spearman correlation of the series with the visit number, and 24 of the 24 flies exhibited positive slopes (p ≤ 0.0001, using sign test).

As illustrated in fig. A1 and summarized in fig. A2 in the Appendix, similar trends are seen in the entry/departure duration ratios.

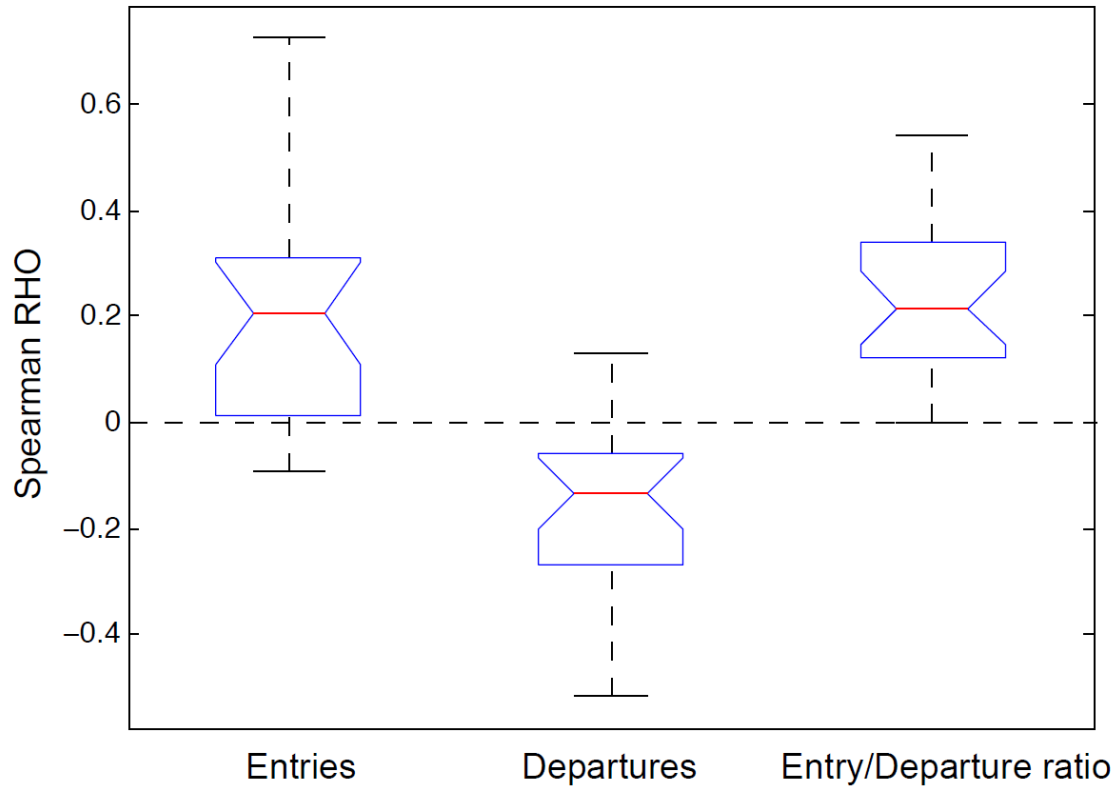

**Fig.8. The developmental dynamics of distance traveled per entry, per departure and per the ratio between successive entry/departure pairs.** As seen by the Spearman RHO the distance traveled per entry increases across the sequence of entries (p=0.0015) while the distance traveled per departure decreases across the sequence of departures (p=0.00028), resulting in an increase across the sequence of entry/departure distance traveled ratios (p<0.0001).

The gradual reduction in the preference for the small arena compared to that of the large arena is thus exhibited in the increased probability of crossing from the small to the large arena, in the decreased probability of crossing from the large to the small arena, and in the increasing distances traveled and time spent in the large compared to the small arena.

The initial reluctance to enter the large arena is illustrated in fig. 1, where before the performance of its first entry, the fly covered ca. 11m during the first 29 hours of the session. Following the first entry into the large arena the fly covered, for another 460 minutes (almost 8 hours), ca. 43 meters (punctuated into entries), before the performance of its first doorway skip (holding back). This two-way phenomenon is illustrated in the animation of the sequence of paths that approach the doorway in the small arena (virtual circle centered on doorway, with radius of 8.5 mm), until the performance of the first substantial entry ([Video A2](#)); and by the sequence of paths in the large arena that approach the doorway and return to the small arena, until performance of the first doorway skipping ([Video A3](#)).

Discussion

*In what sense is the reported behavior neophobic?*

The main objective of this study was to uncover, characterize, and quantify additional novel constraints on fruit fly exploratory behavior. In our experiment a natal fruit was placed in an arena that then became progressively trodden by a freshly eclosed fly, in addition to an untrodden arena, and a distinct, sharp interface between the two arenas, marked by an interruption (doorway) in the wall. The dynamics of behavior exhibited around the interface can be characterized as upstream from well-trodden to untrodden area (relative functional impermeability followed by increased permeability; figure 6 left boxplot), and downstream from untrodden to well-trodden area (almost obligatory crossing followed by optional crossing; figure. 6 right boxplot). At the beginning of the experiment we observed relatively predictable behavior (stereotyped responses), while by the end we observed less predictable responses on both sides of the interface. Finally, there was a progressive increase in relative distance traveled and in relative time spent in the less trodden arena (figures 7, 8, A1, A2). These three dynamic processes imply that the fly's behavior is coupled to how well-trodden the large empty arena is. If the relative time spent in that arena reflects preference, and if the dynamics of crossing reflect a change in preference or, rather, management of the contrast between the two arenas, then the behavior can be described as neophobic. The developmental dynamics of doorway-crossing is the same as that reported in neophobic mice exploring a similar setup [18, 19]. In the fly, coping with neophobia is quantified by the slope depicting the trend in probability of doorway-crossing in the outbound and inbound directions (figures 4, 5). In the mouse, in a similar setup, doorway-crossing is similarly progressive, unfolding gradually during a specific stage of free exploration – the "peep and hide" stage. In the mouse, the buildup was quantified by calculating the portion of the body area (in pixels) extending out of the doorway during peeping [18, 19]). Both mice and flies thus exhibit the same doorway-crossing behavior between well-trodden and untrodden terrain.

*System identification and specification of the system's demand on the CNS*. To study the sensory channels and mechanisms that mediate a behavior it is useful to identify and isolate the behavioral system that is to be studied and to specify the demand it makes on the CNS. We therefore suspended judgment at this stage regarding the sensorimotor mechanisms that mediate the behavior. The operations that appeared to be complex, involving the management of one or more perceptual inputs: visual, olfactory, tactile, and proprioceptive, based on (internal) memory and/or (external) marking behavior. Our present lack of knowledge regarding the actual sensory channels that are used by the fly, and the question of whether the fly's memory is stored internally or deposited externally, does not detract from the demonstration of the very existence of this sensorimotor scheme. For example, staying in the large arena might be mediated by visual memory, by olfactory cues emanating from excretions deposited previously by the fly, by the management of proprioceptive input used for navigation on the basis of self-motion cues [20], or by a combination of several of these sensory inputs. Both mice and flies change the functionality of the doorway by using the same behavior, but this does not imply that they use the same sensorimotor operation. For example, it has been taken for granted that rodents' space neophobia reflects a response to novelty (which implies memory and, therefore, cognition; [18, 19, 21, 22]). If we accept the claim that in the present study fly neophobia has not been proven beyond any doubt to be cognitive, then the same benefit of the doubt should be applied to rodent neophobia. In other words, the space neophobia of either flies or rodents, or of both, might or might not involve cognition. Recently, it has been shown that in a virtual reality setup head fixed fruit flies walking on a ball combine landmark based orientation and angular path integration. When both visual and self-motion cues are absent, a representation of the animal's orientation is maintained in the animal's neural network through persistent activity, a potential substrate for short term memory [20] . Whatever the case, the detailed articulated equivalence of an arthropod and vertebrate space neophobia demands a comparison between the respective neural structures that maintain the behaviors and raises the question of their being homologous.

*Does space neophobia constitute a behavioral homology?* The detailed articulated equivalence between arthropod and vertebrate neophobia could be the result of convergent evolution, or else a genuine Darwinian homology implying common descent. Of necessity, since the argument for common descent cannot be based on fossilized ancestral neophobia, such argument must be based on a shared anatomical and physiological circuitry generating neophobia in both fly and mouse. The circuitry dynamics in the two respective phyla might reflect the circuitry dynamics of a common ancestor. Recently, it has been claimed that a multitude of similarities in network connectivity, embryonic lineages, and genetic programming "suggest deep homology of arthropod central complex and vertebrate basal ganglia circuitries underlying the selection and maintenance of behavioral actions" [23]. Evidence that the central complex is involved in navigation in allocentric space has been reported in several arthropod species [20, 24-26], and the basal ganglia have been reported to promote optimal motor and cognitive control of the environment [27, 28] and map head position in Cartesian space during movement in mice [29].

If the management of untrodden terrain in mice and flies is indeed respectively mediated by homologous underlying structures, then doorway-crossing behavior could be a Darwinian homology. Regardless of whether this behavior is homologous or convergent in the two phyla, the detailed correspondence between them provides an opportunity to use the shared dynamics as a search image for discovering in one phylum features that have already been discovered in the other. For example, the presence of high and low responders to novelty in rodents and the research on the distinct neurochemical characteristics underlying the differences in their behavior [18, 19, 21, 22] suggest that a parallel search in fruit flies might illuminate the mechanisms underlying neophobia in this species.

The progressive increase in the relative occupancy of the less trodden space in flies (Figs. 7, 8, A1,A2 ; video A1) and in mice (movies S1-7 in [18, 19]), and the respective expansion of repertoire in flies (figs 3-7) and in mice (fig. 1 in [18]) suggest that a respective dynamic expansion in space and in behavioral repertoire should be looked for in the neural computations involved in navigation in the two phyla [20, 29].

*Why do we use a piece of fruit in our setup?* In order to demonstrate exploratory abilities, we measured the fruit-fly's unconditioned behavior in a setup that included a strong naturalistic attractor. Such an attractor was provided by a piece of fruit, which is the food substrate for yeast, on which the fly feeds [14, 15]. Fruits induce in *Drosophila* site fidelity, influencing perching locations and heights, foraging distance, and pupation distance from natal fruit, distances moved after disturbance at natal fruit [30], and defense against intruding males [31]. Letting the fly eclose in the vicinity of a piece of fruit was implemented in order to eliminate experimental coercion [13], reduce variability in growing conditions, induce site fidelity, and obtain ethologically relevant exploratory behavior that would be performed in reference to that fruit in the adjacent environment.

*Setup essentials inducing neophobic behavior*. The fruit and the pupa were surrounded by a wall with a wide doorway in it in order to obtain a clear cut interface between natal fruit vicinity and an adjacent, first untrodden, and later relatively untrodden, environment. The natal fruit induced directional polarity across the two arenas. The doorway appeared to both attract fly visits and hinder smooth crossing because of the sharp interface it provided between the well-trodden and untrodden spaces. It may have acted as a one-way perceptual cliff face: while the terrain surrounding the fruit offered a mixture of trodden and untrodden substrate, creating, as it were, a mild gradual increase in how untrodden it was, at the doorway there was a sharp juxtaposition of trodden and untrodden, requiring a discrete outbound "leap" upstream while facilitating the inbound downstream return. The doorway thus functioned as a supernormal stimulus, used in classical ethology for enhancing species-specific behavior to an exaggerated degree, much like the exaggerated brooding response to eggs elicited in oystercatchers by supersize eggs [16]. The doorway also appeared to be a vantage point facing the untrodden, from where the untrodden can be experienced piecemeal (either management of novelty or, for example, management of deposited scent). Repeated visits increased cumulative time spent in the vicinity of the doorway. This was associated with an increase in the versatility of responses to the open doorway, augmenting the probability of either cutting through or holding back (amount of exposure determines repertoire size). The process involved a progressive increase in the relative occupancy of the less trodden arena (video A1).

*Crossing doorways in other species and contexts*. In mice, the cutting through operation is quantified by the extent (frequency and amplitude) of peeping, while holding back is quantified by the frequency of cage skips [18, 19]. Difficulty in doorway crossing ("door phobia") has been reported in dogs, which overcome the problem by crossing the doorway backwards (Video A4). Human Parkinson patients become immobile vis-à-vis doorways by an amount inversely proportional to door width, suggesting a visuo-motor dysfunction [32]. Such dysfunction could not, however, in and of itself, account for functional door impermeability in the flies, because an apparent narrowing of the visual array in the flies impedes crossing only on the way from the trodden to the untrodden. Finally, unimpaired humans showed memory loss after passing through both virtual and real doorways [33, 34].

In vertebrates, spatially distinct compartments are represented by distinct groups of active hippocampal place cells as well as a distinct group of silent cells [35-38]. Since a transition into a novel compartment involves a shift from one set of active and silent cells [35-38] to the generation of a novel set of active and silent cells, which will represent the novel compartment [39], difficulty in transition across doorways could reflect a corresponding difficulty in shifting across sets of place- and silent cells.

Cutting through from the trodden to the untrodden and resisting the attraction of the well-trodden "beaten track" when away from it, are challenging endeavors in both animal and human life. Doorways seem to accentuate the phenomenon by discretizing, and in this way fettering the transition.

*Other operations on the environment:* Examples of invertebrate locomotor operations are walking [40], turning [40], negotiation of barriers and adaptation to slippery ground [41], fixating objects in the face of expanding optic flow [42], and prey capture [43, 44]. Some of the cognitive operations described to date include selecting a specific arm in a Y-Maze [4], selecting a place with reinforcing properties [9, 45, 46], discrimination and generalization learning exhibited by tethered flies [11, 12], habituation in relation to specific stimuli delivered in several sensory modalities in specific behavioral contexts [47], and avoidance of rare unfamiliar flowers in bumble bee pollinators [50]. In vertebrates, operations on the environment include locomotor schemas like that of the mobility gradient [48], and the construction of allocentric space [18, 19, 49]. In allocentric space, a doorway connecting the mouse's home cage with an untrodden arena is used as the origin from which the mouse performs increasingly longer excursions, managing untrodden terrain in a piecemeal, neophobic way.

Operational schemas like the ones described in this study provide a high level vocabulary in robotics for structuring perceptual and motor operations, offering an action-oriented account of behavior and cognition. Avoid obstacle, find intersection, find landmark [50], capture small moving object [51], are but a few examples. The operation "maintain a certain level of novelty" [18, 19, 52] could, for example, account for the progressive gradual increase in exposure to the previously untrodden environment brought about by the repeated probing in "cutting through" and in the progressive increase in staying on the less trodden side in "holding back".

Materials and methods:

*Animals*

We used 24 wild type *Drosophila melanogaster*, 12 males and 12 females of the Oregon strain, raised on a 16:8 light/dark cycle at 25°C. Pupae that were expected to eclose shortly (within a day) were selected and placed in the experimental setup; a single pupa per experiment.

*Experimental Protocol*

Each pupa was placed next to a piece of apple piece (cone shaped, 20mm base diameter and 4mm height) and the fly was then tracked in seclusion, for 72 hours, under constant light and 25°C. We excluded any data collected prior to eclosure.

*Experimental Setup*

The experimental setup consisted of a small (45mm diameter) and a large (130mm) circular arena connected by a 5mm wide open doorway and covered by an 8mm high transparent Perspex ceiling. The flies in our experiment eclosed at their own pace. This dictated long experiments, as it usually took the flies between 10 and 20 hours to eclose and another 16 hours on average to walk the first few meters. [13].

*Videotaping and Tracking*

Each fly was videotaped with a 720X576 pixel resolution camera, 25 frames per second, for a period of 72 hours. The videos were tracked with a MATLAB function called "FTrack", which was developed by Dan Valente [53]. The output received from the "FTrack" function (X and Y coordinates representing the Fly's center of mass) was smoothed using "SEE", a software supported Strategy for the Exploration of Exploration ( http://www.tau.ac.il/~ilan99/see/).

*Defining and scoring the fly's approaches to the doorway*

Using the fly's tracking output, we could easily define the boundaries of the two circular arenas as two ellipses. The midway between the two closest points (where the two ellipses almost "touch" each other) was defined as the doorway's center. Around this center we traced four virtual circles with a radius of 10, 20, 30 and 40 pixels (~3,6,9 and 12 millimeters). The smallest circle had a radius slightly longer than a fly's length and the largest almost reached the fruit, which was placed in the center of the small arena. These circles were used to determine and score the approaches of the fly to the doorway.

Approaches or visits to doorway proximity were then scored as either followed by crossing or by leaving the circle without crossing to the other side (see fig. 2). Having quantified the number of crossings per fixed number of visits for each virtual circle separately we selected the circle and scoring that represented the phenomenon most clearly. Having scored several options we corrected for multiple comparisons using the FDR (False Discovery Rate) to ascertain that we did not have false positives.

*Data analysis using Logistic Regression*

Logistic regression is used to model the probability of crossing the doorway (or not) once at the doorway, as a function of the visit number at doorway. With $p_i$ being the probability of fly $i$ to cross, $p_i/(1-p_i)$ its odds to cross and t the visit number, the model is $log(p_i/(1-p_i))=\beta_0+ \beta t$.

The parameters in the models were estimated using the Matlab function mnrfit (multinomial logistic regression).

For the change in distance traveled the ratios were log-transformed an then the Spearman correlation between these and the number of visit were calculated (no testing was done per fly at this stage because of the obvious serial correlation ).

Testing of the coefficient for both for the group of 24 flies was done using the sign test.

*Fine tuning of arena sizes and fruit location*:

In seeking the optimal parameters that would highlight cognitive behavior we performed preliminary experiments. With two large arenas of the same size and a piece of fruit placed in one of them, the fly did not enter the empty arena a sufficiently large, statistically significant, number of times in the course of three days (a small enough arena increased the likelihood of doorway encounters, which led to entries to the untroden arena). Placing the fruit and the pupa in the small arena or in the large arena yielded an initial preference for the arena where the fruit was placed and a subsequent gradual tendency to cross to the empty arena, thus reducing the asymmetry in the behavior exhibited in the two arenas. Flies that eclosed near a piece of fruit placed in the large arena gradually expanded the maximal distance reached, during successive roundtrips performed from that fruit. We therefore settled on placing the fruit in a small arena connected by a doorway to the large arena.

Acknowledgments:

This study was supported by a grant from the Israel Science Foundation (ISF 1351). We thank Prof. Daniel Segal of the department of Microbiology and Biotechnology, Tel Aviv University, Tel Aviv, Israel, for supplying the experimental animals, and Prof. Segal and Prof. John Ringo of the University of Maine, Orono, ME USA for their advice along the way. Dr. Alex Gomez Marin of Champalimaud Neuroscience Lisbon and Dr. Eyal Gruntman of Janelia Farm HHMI provided highly useful comments on the manuscript. Ms Naomi Paz edited and proofread the manuscript.

46. Wessnitzer, J., Mangan, M., and Webb, B. (2008). Place memory in crickets. Proceedings. Biological sciences / The Royal Society *275*, 915-921.
47. Thompson, R.F., and Spencer, W.A. (1966). Habituation: a model phenomenon for the study of neuronal substrates of behavior. Psychological review *73*, 16.
48. Golani, I. (1992). A mobility gradient in the organization of vertebrate movement: the perception of movement through symbolic language. Behavioral and Brain Sciences *15*, 249-266.
49. Golani, I. (2012). The developmental dynamics of behavioral growth processes in rodent egocentric and allocentric space. Behavioural brain research *231*, 309-316.
50. Arkin, R.C. (1987). Motor schema based navigation for a mobile robot: An approach to programming by behavior. In Robotics and Automation. Proceedings. 1987 IEEE International Conference on, Volume 4. (IEEE), pp. 264-271.
51. Arbib, M.A., and Liaw, J.-S. (1995). Sensorimotor transformations in the worlds of frogs and robots. Artificial Intelligence *72*, 53-79.
52. Gordon, G., Fonio, E., and Ahissar, E. (2014). Learning and control of exploration primitives. Journal of computational neuroscience, 1-22.
53. Valente, D., Golani, I., and Mitra, P.P. (2007). Analysis of the trajectory of Drosophila melanogaster in a circular open field arena. PloS one *2*, e1083.
# Appendix

[Video A1](#) .

Video A1: **Animation of the paths traced during successive entries by a specific fly in the course of a 3 days session**. The current entry is colored in red. History of entries is plotted in gray. Note the progressive increase in extent and complexity of entries. Speed is adjustable by clicking a square brackets key.

[Fly plot, going from the Small arena to the Large arena (Click to view)](#)

Video A2 related to figure 5. **The paths traced by the same selected fly as in video A1 in the familiar arena in the vicinity of the doorway (in pink) until the first entry to the novel arena (in blue).**

[Fly plot, going from the Large arena to the Small arena (click to view)](#)

Video A3 related to figure 6. **All paths traced by the same selected fly as in video A1 upon approaching the doorway from the side of the novel arena**. All paths proceed to the familiar arena (in pink) until the first avoidance (in blue) of the doorway (holding back).

Video [A4](#): **Dogs are reported to have difficulties to cross either a specific doorway or all doorways.** They often cross by walking backwards.

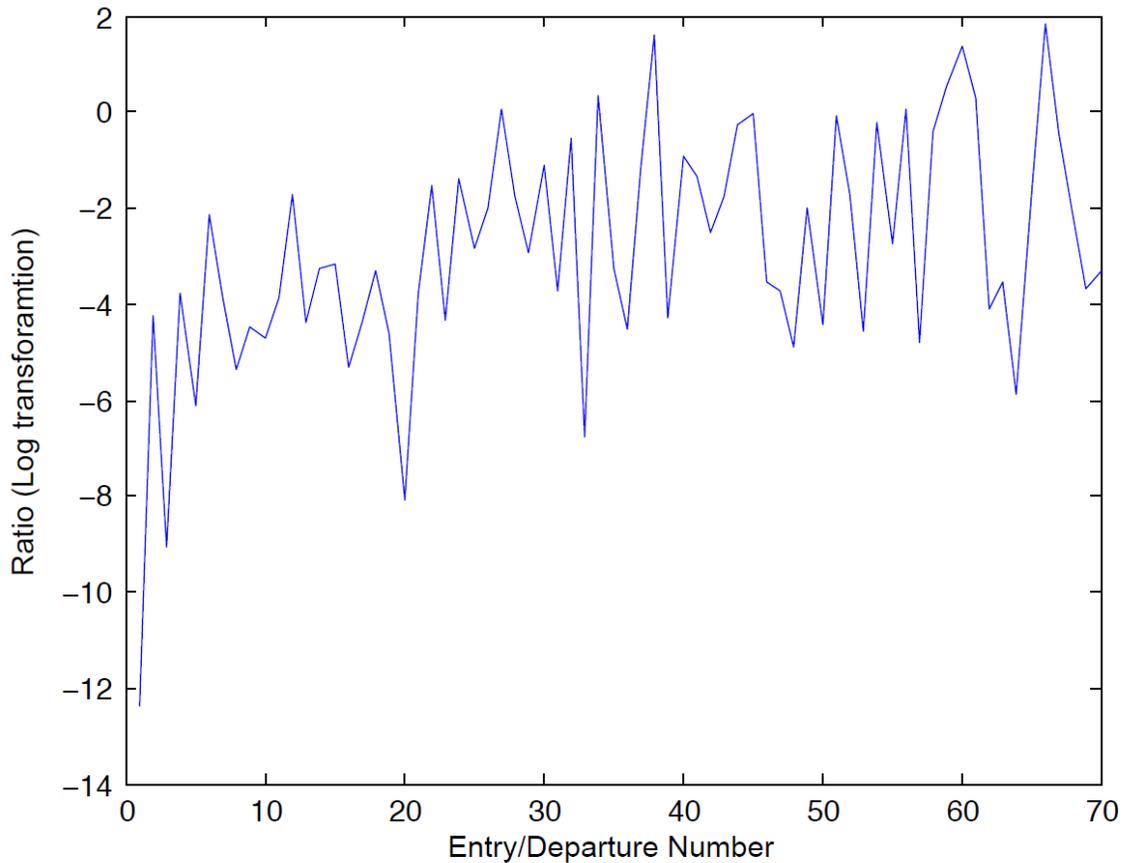

**fig.A1 Related to fig. 7: The sequence of ratios between the duration of each entry and the duration of the departure that preceded it in a selected fly**. This plot also suggests a positive trend, implying a progressive increase in time spent in the large arena relative to time spent in the small arena. A significant positive change in the entry/departure durations ratio is evident in the fly population as a whole (p=0.0015; figure 12 ). The large arena first acts as a repeller and the small arena as an attractor. The repulsive force of the large arena and the attractive force of the small arena are gradually diminished reducing the asymmetry in the behavior in the two compartments.

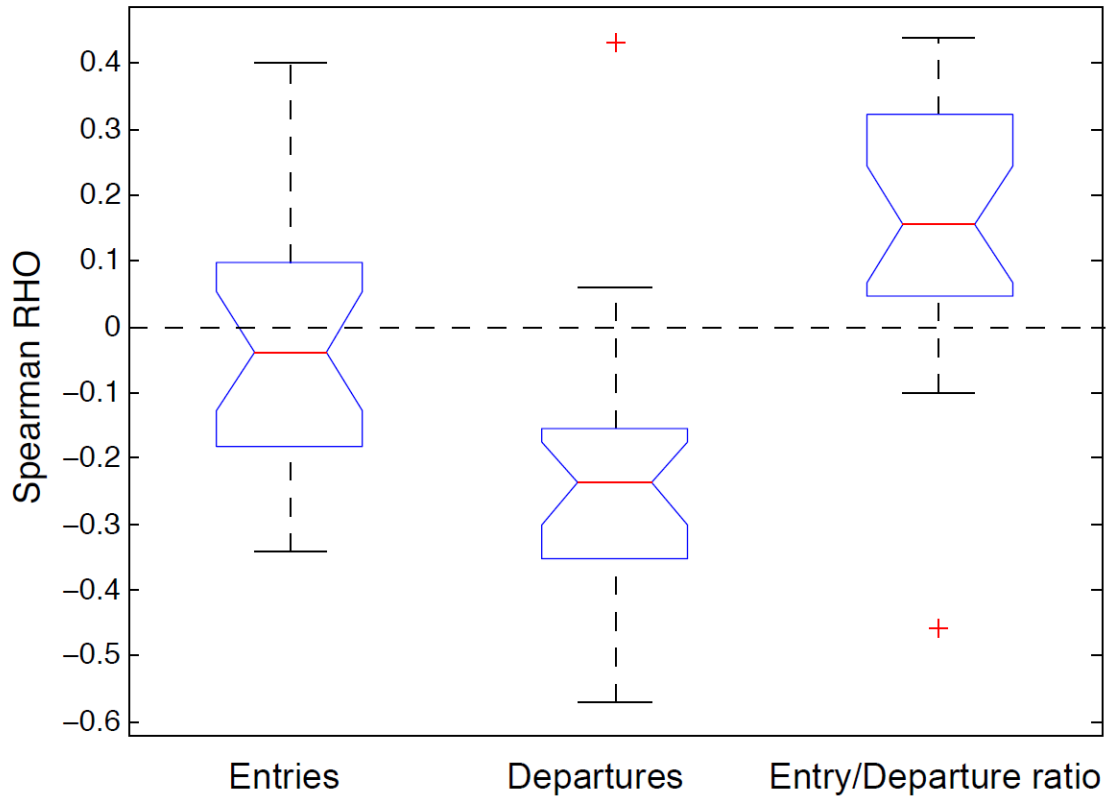

**fig.A2 related to Fig.8: The developmental dynamics of entry duration, departure duration and the ratio between them.** As shown by the Spearman RHO, entry duration does not seem to change across the sequence of entries (p=1) while departure duration decreases across the sequence of departures (p=0.00028), resulting in an increase across the sequence of entry/departure duration ratios (p=0.0015).